\documentclass[amsmath,amssymb,nofootinbib,prd]{revtex4}

\usepackage{bbold}
\usepackage{slashed}
\usepackage{bigints}

\begin{document}

\title{On the Role of Einstein-Cartan Gravity in Fundamental Particle Physics}

\author{Carl F. Diether III}
 \email{fdiether@mailaps.org}

\author{Joy Christian}
 \email{jjc@bu.edu}

\affiliation{Einstein Centre for Local-Realistic Physics, 15 Thackley End, Oxford OX2 6LB, United Kingdom}

\begin{abstract}
Two of the major open questions in particle physics are: (1) Why are the elementary fermionic particles that are so far observed have such low mass-energy compared to the Planck energy scale? And (2), what mechanical energy may be counterbalancing the divergent electrostatic and strong force energies of point-like charged fermions in the vicinity of the Planck scale? In this paper, using a hitherto unrecognized mechanism derived from the non-linear amelioration of Dirac equation known as the Hehl-Datta equation within Einstein-Cartan-Sciama-Kibble (ECSK) extension of general relativity, we present detailed numerical estimates suggesting that the mechanical energy arising from the gravitationally coupled self-interaction in the ECSK theory can address both of these questions in tandem.
\end{abstract}

\maketitle

\parskip 5pt

\baselineskip 14.5pt

\section{Introduction}

For over a century, Einstein's theory of gravity has provided remarkably accurate and precise predictions for the behaviour of macroscopic bodies within our cosmos. For the elementary particles in the quantum realm, however, Einstein--Cartan theory of gravity may be more appropriate, because~it incorporates spinors and associated torsion within a covariant description \cite{Hehl1976,Trautman}. For this reason there has been considerable interest in Einstein--Cartan theory, in the light of the field equations proposed by Sciama \cite{Sciama} and Kibble \cite{Kibble}. For example, in a series of papers Poplawski has argued that Einstein--Cartan--Sciama--Kibble (ECSK) theory of gravity \cite{Hehl-Datta} solves many longstanding problems in physics \cite{Poplawski-1,Poplawski-2,Poplawski-3,Poplawski-4}. His concern has been to avoid singularities endemic in general relativity by proposing that our observed universe is perhaps a black hole within a larger universe \cite{Poplawski-2}. Our concern, on the other hand, is to point out using numerical estimates that ECSK theory also offers solutions to two longstanding problems in particle physics.

The first problem we address here concerns the well known fact that in the limit of vanishing radii (or point limit) the electrostatic and strong force self-energies of point-like fermions become divergent. We will show, however, that torsion contributions within the ECSK theory resolves this difficulty as well, at least numerically, by counterbalancing the divergent electrostatic and strong force energy densities near the Planck scale. In fact, the negative torsion energy associated with the spin angular momentum of elementary fermions may well be the long sought after mechanical energy that counteracts the divergent positive energies stemming from their electrostatic and strong nuclear charges. As a result of this counterbalancing, however, our suggestion does not have anything to do with high energy physics.

The second of these problems can be traced back to the fact that gravity is a considerably weaker ``force'' compared to the other forces. When Newton's gravitational constant is combined with the speed of light and Planck's constant, one arrives at the energy scale of $\sim10^{19}$ GeV, which is some 17 orders of magnitude larger than the heaviest known elementary fermion (the top quark) observed at the mass-energy of ${\sim172}$ GeV. Thus, there is a difference of some 17 orders of magnitude between the electroweak scale and the Planck scale. There have been many attempts to explain this difference, but none is as simple as our explanation based on the torsion contributions within the ECSK theory.

Now one of the reasons why ECSK theory is not widely accepted as a viable theory of gravity is the lack of any experimental evidence for the gravitational torsion. However, as reviewed in \cite{Tecchiolli}, gravitational torsion appears to be entirely confined to elementary fermions, and therefore it is not directly detectable. Elementary fermions are at the heart of all matter and can be viewed as defects in spacetime associated with torsion. Thus, as we have argued elsewhere \cite{diether2}, existence of matter itself may be taken as a proof that gravitational torsion exists, albeit only inside of matter, and therefore it does not propagate. In other words, detection of matter itself may be taken as an indirect detection of gravitational torsion. As we will see, our results below lend  considerable support to this possibility. 

\section{Evaluation of the Charged Fermionic Self-Energy within ECSK theory}

The ECSK theory of gravity is an extension of general relativity allowing spacetime to have torsion in addition to curvature, where torsion is determined by the density of intrinsic angular momentum, reminiscent of the quantum-mechanical spin \cite{Hehl1976,Trautman,Sciama,Kibble,Hehl-Datta,Poplawski-1,Poplawski-2,Poplawski-3,Poplawski-4,Ortin,Rohrlich,Blagojevic,Freidel,Magueijo,Rudenko,Boos}. As in general relativity, the gravitational Lagrangian density in the ECSK theory is proportional to the curvature scalar. Unlike in general relativity, the affine connection
\begin{equation}
\Gamma_{i\,j}^{\;k} = \omega_{i\;\;\;\nu}^{\;\;\mu}e_\mu^k\/e^\nu_j + e^{k}_{\mu}\partial_{i}e^{\mu}_{j}
\end{equation}
is not restricted to be symmetric, although metric compatibility conditions are retained, defined by the requirements that the covariant derivatives of the tetrad fields $e^\mu_k$ and the Minkowski metric vanish, implying the antisymmetry  $\omega_{i}^{\;\;\mu\nu}=-\omega_{i}^{\;\;\nu\mu}$ of the spin connection \cite{fabbri}. The antisymmetric part of the connection, namely 
\begin{equation}
S^k_{\phantom{k}ij}=\Gamma^{\,\,\,\,k}_{[i\,j]}=\partial_i\/e^k_j - \partial_j\/e^k_i
+\omega^{\;\;k}_{i\;\;\;\lambda}e^\lambda_{j}
-\omega^{\;\;k}_{j\;\;\;\lambda}e^\lambda_{i}
\end{equation}
({i.e.}, the torsion tensor), is then regarded as a dynamical variable analogous to the metric tensor $g_{ij}$ in general relativity. Consequently, variation of the total action for the gravitational field and matter with respect to the metric tensor gives Einstein-type field equations that relate the curvature to the dynamical energy-momentum 
tensor $T_{ij}=(2/\sqrt{-g})\delta\mathfrak{L}/\delta g^{ij}$, where $\mathfrak{L}$ is the matter Lagrangian density. On the other hand, variation of the total action with respect to the torsion tensor gives the Cartan equations for the spin tensor of matter \cite{Hehl-Datta}:
\begin{equation}
s^{ijk}=\frac{1}{\kappa} S^{[{ijk}]}\,, \;\;\;\;\text{where}\;\;\kappa\,=\frac{8\pi G}{c^4}.\label{Car}
\end{equation}

Thus, ECSK theory of gravity extends general relativity to include intrinsic spin of matter, with~fermionic fields such as those of quarks and leptons providing natural sources of torsion. Torsion, in turn, modifies the Dirac equation for elementary fermions by adding to it a cubic term in the spinor fields, as observed by Kibble, Hehl and Datta \cite{Hehl1976,Kibble,Hehl-Datta}.

It is this nonlinear Hehl--Datta equation that provides the theoretical background for our proposal. The cubic term in this equation corresponds to an axial-axial self-interaction in the matter Lagrangian, which, among other things, generates a spinor-dependent vacuum-energy term in the energy-momentum tensor (see, for example, Reference \cite{Freidel}). The torsion tensor ${S^k_{\phantom{k}ij}}$ appears in the matter Lagrangian via covariant derivative of a Dirac spinor with respect to the affine connection. The~spin tensor for the Dirac spinor $\psi$ can then be derived \cite{Hehl-Datta}, and turns out to be totally antisymmetric:
\begin{equation}
s^{ijk}= -\frac{i\hbar c}{4}\bar{\psi}\gamma^{[i}\gamma^j\gamma^{k]}\psi, \label{spin}
\end{equation}
where ${\bar{\psi}\equiv \psi^{\dagger}\gamma^0:=(\psi^*_1,\;\psi^*_2,\;-\psi^*_3,\;-\psi^*_4)}$ is the Dirac adjoint of ${\psi}$ and $\gamma^i$ are the Dirac matrices: $\gamma^{(i}\gamma^{j)}=2g^{ij}$. The Cartan Equations (\ref{Car}) render the torsion tensor to be quadratic in spinor fields. Substituting it into the Dirac equation in the Riemann--Cartan spacetime with metric signature ${(+,\;-,\;-,\;-)}$ gives the cubic Hehl--Datta equation \cite{Hehl1976,Kibble,Hehl-Datta}:
\begin{equation}
i\hbar\,\gamma^k\psi_{:k} = mc \, \psi+\frac{3\kappa\hbar^2c}{8}\left(\bar{\psi}\gamma^5\gamma_k\psi\right)\gamma^5\gamma^k\psi, \label{HD}
\end{equation}
where the colon denotes a general-relativistic covariant derivative with respect to the Christoffel symbols, and $m$ is the mass of the spinor. The Hehl--Datta Equation (\ref{HD}) and its adjoint can be obtained by varying the following action with respect to $\bar{\psi}$ and $\psi$, respectively, without varying it with respect to the metric tensor or the torsion tensor \cite{Freidel}:
\begin{equation}
{\cal I} =  \bigintsss d^4x \, \sqrt{-g\,}\left\{-\,\frac{1}{\kappa}\, R + i\hbar c (\bar{\psi} \gamma^k\psi_{:k} - \bar{\psi}_{:k} \gamma^k\psi) - mc^2 \bar{\psi} \psi - \frac{3\kappa \hbar^2 c^2}{8}(\bar{\psi}\gamma^5\gamma_k\psi)(\bar{\psi}\gamma^5\gamma^k\psi)\right\}. \label{action}
\end{equation}

The last term in this action corresponds to the effective axial-axial, self-interaction mentioned~above:
\begin{equation}
\mathfrak{L}_{\rm AA}=-\,\sqrt{-g\,}\,\frac{\,3\kappa \hbar^2 c^2}{8}(\bar{\psi}\gamma^5\gamma_k\psi)(\bar{\psi}\gamma^5\gamma^k\psi). \label{four}
\end{equation}

This self-interaction term is not renormalisable. It is an {effective} Lagrangian density in which only the metric and spinor fields are dynamical variables. The original Lagrangian density for a Dirac field in which the torsion tensor is also a dynamical variable (giving the Hehl--Datta equation), {is} renormalisable, since it is quadratic in spinor fields. As we will see, renormalisation may not be required if ECSK gravity turns out to be what is realised in Nature, because it gives physical justification for the counter terms.

Before proceeding further we note that the above action is not the most general possible action within the present context. In addition to the axial-axial term, an axial-vector and a vector-vector term can be added to the action, albeit as non-minimal couplings (see, for example, Reference \cite{Rudenko}). However, it has been argued in Reference \cite{Freidel} that minimal coupling is the most natural coupling of fermions to gravity because non-minimal couplings are sourced by components of the torsion that do not appear naturally in the models of spinning matter. For this reason we will confine our treatment to the minimal coupling of fermions to gravity and the corresponding Hehl--Datta equation, while~recognizing that strictly speaking our neglect of non-minimal couplings amounts to an approximation, albeit a rather good approximation, at least as far as electrodynamics is concerned.

\subsection{S-Matrix Evaluation of the Charged Fermionic Self-Energy within QED} \label{2a}

It is instructive for our purposes to first review in this subsection the standard treatment of the charged fermionic self-energy within QED, ignoring the ECSK gravity. To this end, recall that the canonical evaluation of the electron self-energy was performed by Weisskopf in 1939 \cite{Weisskopf}, which~revealed that in general the self-energy of an electron is logarithmically divergent. An S-matrix evaluation for the electromagnetic mass, $\delta m$, confirms Weisskopf's result \cite{Milonni,Weinberg,Halzen}. In the units of $\hbar = c = 1$ and $\alpha = e^2/4\pi$, together with a high-energy cutoff ${\Lambda}$, this electromagnetic mass can be expressed as 
\begin{equation}
\delta m = \frac{3 \alpha \,m}{2\pi}\ln{\left(\frac{\Lambda}{m}\right)}\,\,\,\,(\text{for}\; \Lambda >> m), \label{milonni1}
\end{equation}
where ``ln'' stands for the natural logarithm.  Let us investigate this result using simple numerical analysis. What happens, for example, if we use Planck mass for the cutoff?  This should be a reasonable assumption since the Planck scale can be considered a limiting scale in the same fashion as $\hbar$ and $c$ are considered. Then, for an electron we have
\begin{equation}
\delta m_e = \frac{3 \alpha \,m_e}{2\pi}\ln{\left(\frac{m_P}{m_e}\right)} \approx 0.1795 \,m_e. \label{selfenergy1}
\end{equation}

This suggests that the electromagnetic mass-energy of an electron in this case would be about 18~percent of the total rest mass-energy.  That does not seem unreasonable, but there is no way to confirm that it is correct. Moreover, according to some views, if the electromagnetic mass is truly infinite then there should be a compensating negative infinite mechanical mass that produces the observed positive rest mass. On the other hand, Weinberg in Reference \cite{Weinberg} gives a renormalised expression for the ``complete self-energy function'' as
\begin{align}
\Sigma^*_{order \,e^2} (p) &= \Sigma^*_{1\,loop}(p)-(Z_2-1) (i \slashed{p} + m_x)+Z_2\,\delta m_x \nonumber \\
&=\frac{-2\pi^2 e^2}{(2\pi)^4}\bigintss_{0}^{1} dx \, \Bigg\{ [i(1-x) \slashed{p}+2 m_x] \ln\left({\frac{m_x^2(1-x)}{p^2 x(1-x)+m_x^2 x}}\right) \nonumber \\
& \;\;\;\;\;\;\;\;\;\;\;\;\;\;\;\;\;\;\;\;\;\;\;\;\;\;\;\;\;\,\quad
- m_x[1+x]\ln\left({\frac{1-x}{x^2}}\right) \nonumber \\
& \;\;\;\;\;\;\;\;\;\;\;\;\;\;\;\;\;\;\;\;\;\;\;\;\;\;\;\;\;\;\;\;\;\;\,\quad -(i\slashed{p}+m_x)\left[(1-x)\ln\left({\frac{1-x}{x^2}}\right)- \frac{2(1-x^2)}{x}\right] \Bigg\}. \label{rcsef}
\end{align}

The middle term in this integrand represents the renormalised electromagnetic mass so that for an electron it gives 
\begin{equation}
\delta m_{e\,renorm} = \frac{-2\pi^2 e^2}{(2\pi)^4}\bigintsss_{0}^{1} dx \left\{-m_x[1+x]\ln\left({\frac{1-x}{x^2}}\right)\right\} =\frac{3 \alpha \,m_e}{8\pi} \approx 0.0008711 \,m_e. \label{selfenergy2} 
\end{equation}

In this case the electromagnetic mass is considerably less than 1 percent of the total rest mass. That indicates that the cutoff should be very close to the rest mass of the electron.  Thus we seem to have two conflicting results.  The conflict can be resolved by realizing that the mass in the equations leading to these results was plugged in by hand in an {ad hoc} manner. Moreover, from a quantum field theory and Feynman diagram perspective, it is easy to see that the mass comes in via the fermion propagator. Therefore by definition it is off-mass-shell, since the propagator represents an internal virtual line.  It can thus have any value. In other words, it is really a variable. For the time being we will treat it as a variable and use the relation $m = 1/r$.  Now that the apparent conflict is resolved, in~what follows we focus on the renormalisation process.

It is easy to see that, in the rest frame of the particle in question with $i\slashed{p}=-m$, the renormalised complete self-energy function in Equation (\ref{rcsef}) is identically zero.  That may seem odd, since one would expect that in the rest frame the self-energy would be equal to the rest mass-energy. It really just means that the propagator is zero, as it should be for the rest frame. Thanks to the Hehl--Datta equation, we~have an additional term in the Lagrangian to consider in the renormalisation process.  In our view, this additional term represents the mechanical (bare) mass, since it is negative with respects to the observed mass.  Thus, in the renormalised self-energy function in Equation (\ref{rcsef}), the counter terms are superfluous since the torsion term from the Hehl--Data equation is the counter term! Moreover, we~will see later that the rest mass is produced in a natural way from energy considerations generated only from the physical constants and geometry.

Using $m = 1/r$, we can now put the renormalised electromagnetic mass equation in a more familiar form for electromagnetic energy,
\begin{equation}
\delta m_{renorm} = \frac{3 \alpha m}{8\pi} = \frac{3\alpha}{8\pi r}  = \frac{3\, e^2}{32\pi^2 \,r}. \label{milonni2}
\end{equation}

However, $\delta m$ in this expression still goes to infinity as $r \rightarrow 0$. Perhaps a Planck length cutoff may be used to tame this infinity. We will soon see that $r$ does take on a finite value very close to Planck length. Now we know from experiments that the radius of an electron is likely to be less than \mbox{$10^{-22}$ m \cite{Dehmelt}}. Substituting that bound for $r$ gives
\begin{equation}
\delta m = \frac{3\, e^2}{32\pi^2\,(10^{-22}\textnormal{m})} \gtrapprox 1.719 \times 10^{3}\, \textnormal{GeV}, 
\end{equation}
which, although finite, is still a very large electromagnetic energy contribution, with the actual value likely to be even greater. Thus, according to this estimate, the electromagnetic mass is going to be very large near the Planck length and will have to be compensated for in order to recover the observed rest mass of a charged fermion. The compensation will have to be negative mass-energy relative to the positive electromagnetic energy:  
\begin{equation}
m_{obs} = -X + \delta m_R = -X + \frac{3\,\alpha}{8\pi r_B} \longrightarrow -\frac{X}{r} + \frac{3\,\alpha}{8\pi r_B}. \label{milonni4}
\end{equation}

We suspect that the unknown variable $X$ might be related to the Hehl--Datta self-interaction term~(\ref{four}) because that term varies as $1/r_x$. Our goal then is to investigate this possibility.

To that end, note that the full second order S-matrix calculation within QED worked out by Milonni \cite{Milonni} gives
\begin{align}
S_{fi}^{(2)}(E) &= -i(2\pi)^4 \delta^4(p_f - p_i)\sqrt{\frac{m^2}{E_i E_f}}\frac{1}{r^3}\bar{u}(p_f, s_f)\Sigma(p_i)u(p_i, s_i), \label{milonni5} \\ 
\noindent\text{where}\;\;\;
\Sigma(p_i) &= -i e^2 \bigintssss{\frac{d^4 k}{(2\pi)^4}\frac{g_{\mu\nu}}{k^2+i\epsilon}\gamma^{\mu}\frac{1}{\slashed{p}_i - \slashed{k} - m + i\epsilon}\gamma^{\nu}}.  \label{milonni5a}
\end{align}

Milonni's evaluation of $\Sigma(p_i)$ produces the result for $\delta m$ given in Equation (\ref{milonni1}). However, we doubt this evaluation is appropriate since, as we noted previously, the mass comes in via the propagator and therefore it is off mass shell.  We will not use it in our evaluation.

\subsection{S-Matrix Evaluation for the Charged Fermionic Self-Energy within Einstein--Cartan Gravity}

Building on the results of the previous subsection, we now evaluate the self-energy S-matrix process including the contribution from ECSK gravity. Our calculation will be different from the old way of calculating self-energy in the manner of Weisskopf because in our approach we have the gravitational torsion term as a built-in counter-term. The following calculations will be somewhat unconventional as there is no Feynman diagram for evaluation in the rest frame.  We begin with our complete QED Lagrangian for free particles:
\begin{equation}
\frac{\mathfrak{L}_{\rm QED}}{\sqrt{-g\,}}= i\hbar c\,\bar{\psi} \gamma^{\mu}\partial_{\mu}\psi + e\,\bar{\psi}\gamma^{\mu}A_{\mu}\psi - \frac{1}{4} F_{\mu\nu}F^{\mu\nu} - mc^2 \bar{\psi} \psi - \frac{3\kappa \hbar^2 c^2}{8}(\bar{\psi}\gamma^5\gamma_{\mu}\psi)(\bar{\psi}\gamma^5\gamma^{\mu}\psi).
\end{equation}

Here the last term is from the Hehl--Datta equation for gravitational torsion via spin density squared. We take it to be a self-interaction involving intrinsic spin.  In other words, the spin interacts with itself much like the charge interacts with the field it creates.  For our analysis of a charged lepton in the rest frame, only three terms are applicable because the derivative term is zero and normal gravity is negligible.  The self-interaction Hamiltonian density for the first term with $\hbar = c =1$ and $(x) =({\bf x}, t)$ is then
\begin{equation}
\textnormal{h}_I(x)_1 = e\,\bar{\psi}(x)\gamma^{\mu}A_{\mu}\psi(x). \label{ED}
\end{equation}

The next self-interaction term is
\begin{equation}
\textnormal{h}_I(x)_2 = - \frac{\,3\kappa}{8}(\bar{\psi}(x)\gamma^5\gamma_{k}\psi(x))(\bar{\psi}(x)\gamma^5\gamma^{k}\psi(x)). \label{TD} 
\end{equation}

For our purposes it is sufficient to evaluate the S-matrix in the rest frame, where the quantities $\bar{\psi}(x)\gamma^5\gamma_{k} \psi(x)$ and $\bar{\psi}(x)\gamma^5\gamma^{k} \psi(x)$ appearing on the RHS are nonzero even in the rest frame because the coupling between fermions and anti-fermions is an intermediate step in the self-interaction, as~we have shown in Equations (\ref{s-matrix0}) and (\ref{s-matrix_anti}) of Appendix \ref{B}. Furthermore, as shown in Appendix \ref{B}, the~first order S-matrix terms in the rest frame are now simplified to
\begin{equation}
S_{fi}^{(1)}(E1) = -it\frac{\alpha}{r}
\left(\begin{array}{cccc}
1& 0& 0& 0\\
0& 1& 0& 0\\
0& 0& 0& 0\\
0& 0& 0& 0
\end{array}\right). \label{ED2}
\end{equation}
\begin{equation}
S_{fi}^{(1)}(E2) = it\frac{\,3\kappa}{8\,r^3}
\left(\begin{array}{cccc}
1& 0& 0& 0\\
0& 1& 0& 0\\
0& 0& 0& 0\\
0& 0& 0& 0
\end{array}\right). \label{TD2}
\end{equation}

On the other hand, for $m\bar{\psi}\psi$ in the rest frame, applying Equation (\ref{b5}) from Appendix \ref{B} we have
\begin{equation}
S_{fi}^{(1)}(E3)= i m \int{d^4x\, \langle f\mid \bar{\psi}(x)\psi(x)\mid i \rangle} = i\frac{m\,r^3\,t}{r^3}\,\bar{u}^f(m) {u}^i(m) = i\,m t
\left(\begin{array}{cccc}
1& 0& 0& 0\\
0& 1& 0& 0\\
0& 0& 0& 0\\
0& 0& 0& 0
\end{array}\right).
\end{equation}

Equation (\ref{HD}) suggests that the sum of these three terms is equal to zero:
\begin{align}
   - it\frac{\alpha}{r}
\left(\begin{array}{cccc}
1& 0& 0& 0\\
0& 1& 0& 0\\
0& 0& 0& 0\\
0& 0& 0& 0
\end{array}\right)+it\frac{\,3\kappa}{8\,r^3}
\left(\begin{array}{cccc}
1& 0& 0& 0\\
0& 1& 0& 0\\
0& 0& 0& 0\\
0& 0& 0& 0
\end{array}\right)
+ i\,m t
\left(\begin{array}{cccc}
1& 0& 0& 0\\
0& 1& 0& 0\\
0& 0& 0& 0\\
0& 0& 0& 0
\end{array}\right) &= 0.
\end{align}

Consequently, we have
\begin{equation}
\frac{\alpha}{r} - \frac{\,3\kappa}{8\,r^3} = m.
\end{equation}

Reinstating $\hbar$ and $c$ we thus arrive at
\begin{equation}
\frac{\alpha \hbar c}{ r} - \frac{\,3\kappa (\hbar c)^2}{8\,r^3} = m c^2. \label{result}
\end{equation}

An identical result can be obtained also for the anti-fermion spinor, $v(m)$, so that this equation remains the same for both fermion and anti-fermion.  In what follows we will use this S-matrix result for our numerical approximations.  From these results, after solving for $r$, it is evident that the complete process is finite, without divergences. This suggests that the above is the correct S-matrix solution for the fermion self-energy problem, with all other orders of the S-matrix expansion vanishing because the {genuine} fermionic self-energy must naturally be evaluated only in the rest frame, with all other contributions summing to zero. Any higher loop corrections in squared charge will be automatically compensated by higher loop corrections from the squared spin.

It is also worth noting that without ameliorating the Dirac equation with a cubic term, the Dirac equation would reduce for an electron to $\alpha \hbar/r_e=m_ec$, giving $r_e=\alpha \hbar/(m_ec)\sim 10^{-15}$ m, where~$\alpha = e^2/(4\pi\hbar c)$ is the fine structure constant. This is the classical electron radius. Experimental evidence, however, suggests that electron radius is much smaller \cite{Dehmelt}. As we shall see, our calculations with the cubic term included predicts the electron radius to be of the order of ${10^{-34}}$ m, which is closer to the Planck length. This may turn out to be the correct value of the electron radius.

Needless to say, what we have presented above is a derivation of Equation (\ref{result}) within a theory that may be viewed as a quantum field theory of Dirac fields in a Riemann--Cartan spacetime \cite{Kibble,Hehl-Datta}. It~can be interpreted also as a theory of gravity-coupled self-interaction within standard general relativity \cite{Hehl1976,Hehl-Datta}. However, any such generalization must necessarily reproduce the Hehl--Datta Equation~(\ref{HD}) for single fermions even at reasonably high energies, just as Dirac equation remains valid for single fermions at high energies \cite{Boos}.

Finally, it is important to note here that, despite the appearance of four spinors in the interaction term of Equation (\ref{action}), it describes the self-interaction of a {single} fermion, of range ${\sim 10^{-34}}$ m, not~mutual interactions among the spins of four distinct fermions. That is to say, it does not describe a ``spin field'' of some sort as a carrier of a new interaction \cite{Hehl-Datta}. If, however, one insists on interpreting the interaction term in Equation (\ref{action}) as describing interactions among four distinct fermions, then the mass of the corresponding exchange boson would have to exceed ${10^{15}}$ GeV, which is evidently quite unreasonable. Moreover, as we shall see in Section \ref{III}, this energy is a fictitious quantity and therefore there is no justification for assuming some kind of a new torsion interaction between different fermions. What~is more, as we shall soon see, within our scheme any corrections due to vacuum polarization are automatically compensated for in the production of electroweak mass-energy, dictated by \mbox{Equation~(\ref{result}) above}.

\subsection{Evaluation of Charged Fermionic Self-Energy by Dimensional Analysis}

It is instructive to compare the results of the previous two subsections with the evaluation of the charged fermionic self-energy using only dimensional analysis. To this end, we begin with the following physically reasonable assumptions:
\begin{enumerate}
\item ECSK theory of gravity is the correct theory of spacetime for addressing the fermionic self-energy problem since it allows the dimension-full gravitational constant, $G$, to enter elementary particle physics in a natural manner.

\item Since experiments to date indicate that an electron is a point-like particle without any substructure and put an upper bound of $10^{-22}$ meters on its radius, we assume that the radius of electron is much less than that value. 

\item We assume that the radial distance, $r$, on which the electromagnetic self-interaction depends is the same radial distance on which the self-interaction arising from the ECSK gravity-induced torsion spin density also depends.
\end{enumerate}

Given these assumptions, we ask: What physical mechanism is responsible for the observed rest mass $m_x$ of the elementary charged fermions? To answer this question we express the rest mass energy in CGS units, and assume that it is at least partially\footnote{As is well known, assuming that the rest mass energy is entirely electromagnetic in nature leads to the classical radius of the electron, which has been ruled out by experiments. Our assumption of it being at least partially electromagnetic in nature is quite reasonable.} electromagnetic in nature, so that it satisfies a relation like
\begin{equation}
m_x c^2 \sim \frac{e^2}{r} + X,
\end{equation}
where the dimensionality of $X$ is necessarily that of energy.  However we already know that the value of $r< 10^{-15}\,m$ produces an energy greater than $m_x c^2$. Therefore $X$ must be negative energy, giving
\begin{equation}
m_x c^2 \sim \frac{e^2}{r} + (-X).
\end{equation}

Now, since fermions have spin $\hbar/2$, it is reasonable to assume that it is involved in a mechanical-like energy resulting from the spin interacting with itself analogous to charge, so that we may have a relation like
\begin{equation}
-X \sim -\left(\frac{\hbar}{2}\right)^2 \frac{1}{r}.
\end{equation}

It is evident from this expression that what we have on its RHS is energy $\times$ mass $\times$ length $= E M L$, so we will have to divide out by mass and length to get the dimensions of energy, giving
\begin{equation}
-X \sim -\left(\frac{\hbar}{2 r}\right)^2 \frac{1}{M},
\end{equation}
which in terms of the gravitational constant ${G\sim L^3/MT^2}$ can be written as
\begin{equation}
-X \sim -G \left(\frac{\hbar}{2 r}\right)^2.
\end{equation}

The dimensions on the RHS of this expression now give us $E\times L^3/T^2$, so we will have to cancel out $L^3/T^2$. A natural candidate to accomplish that is the speed of light, $c$, giving
\begin{equation}
-X \sim -\frac{G}{c^2 r} \left(\frac{\hbar}{2 r}\right)^2 = -\frac{G}{r^3} \left(\frac{\hbar}{2 c}\right)^2,
\end{equation}
provided we cancel out the extra $L$ with $1/r$. Thus, we now have the dimensions of energy on the RHS, so that we finally have
\begin{equation}
m_x c^2 \sim \frac{e^2}{r_x} - \frac{G}{r_x^3} \left(\frac{\hbar}{2 c}\right)^2.
\end{equation}

This is the basic form of our central equation (apart from some numerical constants) which we have derived also using two other methods in the previous subsections. Solving this  equation for $r_x$ with electron mass for $m_x$ gives a value of the order of $10^{-34}\, $m. However, rather surprisingly, we also found a solution for $r_x$ for the classical electron radius.

\section{Particle Masses via Torsion Energy Contribution}\label{III}

We start with our S-matrix result for a charged fermion:
\begin{equation}
\frac{\alpha \hbar c}{r} - \frac{\,3\pi G \hbar^2}{c^2\,r^3} = m c^2. \label{result3}
\end{equation}

The first term in Equation (\ref{result3}) diverges as ${r \rightarrow 0}$. If we set $r$ to Planck length we obtain 
\begin{equation}
\frac{\alpha \hbar c}{l_P} \approx 8.909 \times 10^{16}\, \textnormal{GeV}, \label{pl}
\end{equation}
which is close to Planck energy. Although finite, this is still an extremely large energy. {Such a large energy for charged leptons is never realised in Nature}. A natural question then is: Is there a negative mechanical energy that cancels out most of this energy to produce the observed rest mass-energy of leptons? We believe the answer lies in the second term of Equation (\ref{result3}), which---as we saw above---arises from the non-linear amelioration of the Dirac equation within ECSK theory. Indeed, if~we again set the Planck length for $r$ in the second term of Equation (\ref{result3}), then we obtain 
\begin{equation}
-\frac{\;3\pi\,G\hbar^2}{2\,c^{2}\,(l_{P})^{3}} \approx -\,5.773\times 10^{19}\ \textnormal{GeV}. \label{meng}
\end{equation}

Comparing this value with the electrostatic energy at the Planck length estimated in Equation (\ref{pl}) we see at once that the torsion-induced mechanical energy in Equation (\ref{meng}) can indeed counterbalance the huge electrostatic energy. This is a surprising observation, considering the widespread belief that ``the numerical differences which arise [between GR and ECSK theories] are normally very small, so~that the advantages of including torsion are entirely theoretical'' \cite{Boos}.

Moving forward to our goal of numerical estimates, let us note that whenever terms quadratic in spin happen to be negligible, then the ECSK theory is observationally indistinguishable from general relativity. Therefore, for post-general-relativistic effects, the density of spin-squared has to be comparable to the density of mass. The corresponding characteristic length scale, say for a nucleon, is~referred to as the Cartan or Einstein--Cartan radius \cite{Trautman,Boos}, defined as
\begin{equation}
r_{Cart} \approx (l_P^2\,\lambda_{C})^{\frac{1}{3}}, \label{cart}
\end{equation}
where ${\lambda_{C}}$ is the Compton wavelength of the nucleon. Now it has been noted by Poplawski \cite{Poplawski-1,Poplawski-2,Poplawski-3,Poplawski-4,Poplawski-5,Poplawski-6} that quantum field theory based on the Hehl--Datta equation may avert divergent integrals normally encountered in calculating radiative corrections, by self-regulating propagators. Moreover, the multipole expansion applied to Dirac fields within the ECSK theory shows that such fields cannot form singular, point-like configurations because these configurations would violate the conservation law for the spin density, and thus the Bianchi identities. These fields in fact describe non-singular particles whose spatial dimensions are at least of the order of their Cartan radii, defined by the condition
\begin{equation}
\epsilon\sim\kappa s^2,
\end{equation}
where ${\sqrt{s^2}\sim \hbar c\,|\psi|^2}$ is the spin density, ${\epsilon\sim mc^2|\psi|^2}$ is the rest energy density and ${|\psi|^2\sim 1/r^3}$ is the probability density, giving the radius in Equation (\ref{cart}). Consequently, at the least the de Broglie energy associated with the Cartan radius of a fermion (which is approximately ${10^{-27}\,}$m for an electron) may introduce an effective ultraviolet cutoff for it in quantum field theory in the ECKS spacetime.  The~avoidance of divergences in radiative corrections in quantum fields may thus come from spacetime torsion originating from intrinsic spin. Poplawski and others, however, took $\epsilon$ to be the mass-energy density of the fermion to arrive at the Cartan radius Equation (\ref{cart}). It is easy to work out from the first term of our Equation (\ref{result3}) that at the Cartan radius the electrostatic energy density for an electron is still extremely large:
\begin{equation}
\frac{\alpha \hbar c}{(10^{-27}m)^4} \approx 1.440 \times 10^{90}\, \textnormal{GeV\,m}^{-3}.
\end{equation}

For this reason it is not correct to identify $\epsilon$ with the rest mass-energy density, which is ${\approx 5.1099 \times 10^{77}\ \textnormal{GeV}\  \textnormal{m}^{-3}}$ for an electron at the Cartan radius. The electrostatic energy density of an electron is thus about thirteen orders of magnitude higher. Therefore $\epsilon$ is better identified with the electrostatic energy density provided most of it is cancelled out. 

If in Equation (\ref{result3}) we set the electrostatic energy appearing in its first term to be equal to the spin squared energy induced by the self-interaction appearing in its second term and solve for ${r}$, then we~obtain
\begin{equation}
\boxed{r_t = \sqrt{{\frac{3\pi}{\alpha\,}}}\;l_P}\,,
\end{equation}
the value of which works out to be
\begin{equation}
r_t \approx 5.808 \times 10^{-34}\, \rm{m}\,.
\end{equation}

Thus, our ${r_t}$ is about 36 times larger than the Planck length, and, as we can see, it is a remarkably simple constant in terms of the Planck length, $l_P$, and the fine structure constant, $\alpha$.  According to Equation (\ref{result3}), $r_t$ is the effective radius at which energy density due to spin density should completely compensate the huge electrostatic energy seen in Equation (\ref{pl}). In our view, this is the correct Cartan radius, at least for the charged leptons, and that may still provide a plausible mechanism for averting singularities, since it is still larger than the Planck length. {It is important to note, however, that these huge energy densities never actually occur in Nature, because according to our Equation (\ref{result3}) they are automatically compensated}. The physical mechanism described above is simply to enable extraction of the radius ${r_x}$ for different charged fermions.

We can now use Equation (\ref{result3}) to solve for $r_x$ for the different charged leptons and anti-leptons which leads us to the following formula for our numerical estimates:
\begin{equation}
\frac{\alpha \hbar c}{r_{x}} - \frac{3\, \pi G \hbar^2}{c^2\, r_x^3} = +m_{x}c^{2}\,. \label{e}
\end{equation}

As shown in the Appendix \ref{A} below, we were able to find solutions for $r_x$ for the charged leptons using arbitrary precision in {\sl Mathematica}. The first in our results listed below is the solution for $r_t$ up to 24 significant figures. Then, using the same precision for comparison, we list the results for $r_e$ for an electron, $r_{\mu}$ for a muon and $r_{\tau}$ for a tauon, along with the anti-fermions:
\begin{align*}
r_t\,\,\,\, &= 5.80838808109165274355010 \times 10^{-34}\ \textnormal{m} \longrightarrow 0.0  \textnormal{\ MeV}\,, \\
r_{e-} = r_{e+} &= 5.80838808109165274414872 \times 10^{-34}\  \textnormal{m} \longrightarrow 0.511  \textnormal{\ MeV}\,, \\
r_{\mu-} = r_{\mu+} &= 5.80838808109165286732523 \times 10^{-34}\  \textnormal{m}\longrightarrow 106  \textnormal{\ MeV}\,, \\
r_{\tau-} = r_{\tau+} &= 5.80838808109165482503366 \times 10^{-34}\  \textnormal{m}\longrightarrow 1777  \textnormal{\ MeV}\,.
\end{align*}

We should note that there is also a positive solution obtained for the reduced Compton wavelength for these fermions in the form of
\begin{equation}
r_x = \alpha\frac{\hbar}{m_x c}. 
\end{equation}

This is so because the spin density squared term for an electron becomes very small of order $10^{-38}$~MeV when $r_x$ is equal to that result so it can be considered effectively as zero.  This solution for fermion radius appears to have been ruled out by scattering experiments. However, while it is true that no structure has been found via scattering experiments, there does seem to be some structure involving the magnetic moment and zitterbewegung. The apparent conflict with the scattering experiments can be resolved if there is a very small object for electric charge near Planck length, as our solution indicates, that is ``circulating'' about the Compton wavelength. Then the scattering at high energies can be understood as from that point-like object and the scattering at low energies can be understood as from the Compton wavelength ``size'' due to the Coulomb field, with the Coulomb field ``barrier'' near the Compton wavelength penetrated in the high energy scatterings. 

Evidently, very minute changes in the radii are seen to cause large changes in the observed rest mass-energies of the fermions. As the differences in the radii go larger, the resultant mass-energies go higher, as one would expect. It seems extraordinary that Nature would subscribe to such tiny differences resulting from a large number of significant figures, but that might explain why the underlying relationship between the observed values of the masses of the elementary particles has remained elusive so far. In addition to the possible reasons for this mentioned above, it is not inconceivable that the difference between the spin energy density and the electrostatic energy density radii with respect to $r_t$ arises due to purely geometrical factors. We also suspect that there may possibly be some kind of symmetry breaking mechanism at work similar to the Higgs mechanism, and this symmetry breaking results in the observed mass-energy generation.

As a consistency check, let us verify that the tiny length differences seen above vanish, ${\Delta r\rightarrow 0}$, as the corresponding rest mass-energy differences tend to zero: ${\Delta E \rightarrow 0}$. To this end, we recast Equation~(\ref{e}) for arbitrary ${r_x}$ in a form involving only rest mass-energy on the RHS as:
\begin{equation}
\frac{\alpha \hbar c}{r_{x}} - \frac{3 \pi\, l_P^2 \hbar c}{r_x^3} = {m_{x}c^{2}}. \label{rx}
\end{equation}

If we now set
\begin{equation}
A \equiv \alpha \hbar c\;\;\;\;\text{and}\;\;\; B \equiv 3 \pi \, l_P^2 \hbar c\,, \label{A-B}
\end{equation}
then, with ${\Delta E = m_x c^2}$ and setting ${r_x=r_t}$ as the cancellation radius for which ${\Delta E = 0}$, we obtain
\begin{equation}
r_t = \sqrt{B/A}\,. \label{ba}
\end{equation}

This allows us to derive a general expression for ${r_x}$ when ${\Delta E\not= 0}$:
\begin{equation}
\frac{A}{r_{x}} - \frac{B}{r_x^3} = \Delta E. \label{rx2}
\end{equation}

From this expression it is now easy to see that
\begin{equation}
\lim_{{\Delta E}\rightarrow\,0}\left\{\left(\frac{A}{r_{x}} - \frac{B}{r_x^3}\right)  = \Delta E\right\} = \left(\frac{\sqrt{A}^3}{\sqrt{B}} - \frac{\sqrt{A}^3}{\sqrt{B}}\right) \ln{\frac{\sqrt{B/A}}{l_P}} = 0 \implies r_x =\, \sqrt{B/A} \,=\, r_t\,, \label{rx3}
\end{equation}
and conversely, using Equation (\ref{ba}), 
\begin{equation}
\lim_{r_x\,\rightarrow\,r_t}\left\{\left(\frac{A}{r_{x}} - \frac{B}{r_x^{3}}\right) = \Delta E\right\} \implies \Delta E = 0\,. \label{rx4}
\end{equation}

Consequently, with ${\Delta r \equiv |\,r_t - r_x|}$, we see from the above limits that $\Delta r \rightarrow 0$ as $\Delta E \rightarrow 0$, and vice~versa.

As a rough estimate, the calculation for the radius $r_q$ of elementary quarks can be performed in a similar manner as that for charged leptons, since at such short distances the strong force reduces to a Coulomb-like force. One must also factor-in the electrostatic energy, so that a relationship like the following must be calculated, say, for the top quark:
\begin{equation}
\frac{9\,\alpha \hbar c}{4 \, r_{qx}} + \frac{\alpha_s \hbar c}{3 \, r_{qx}} - \frac{3 \pi G \hbar^2}{c^{2} \,r_{qx}^{3}} = m_t c^2\,.  \label{qcal}
\end{equation}

Here $\alpha_s$ is the appropriate strong force coupling (we use $0.1$). Needless to say, a cancellation radius different from that of the charged leptons should be calculated for comparison, by setting
\begin{equation}
\frac{9\,\alpha \hbar c}{4 \, r_{qt}} + \frac{\alpha_s \hbar c}{3 \, r_{qt}} - \frac{3 \pi G \hbar^2}{c^{2} \, r_{qt}^{3}} = 0. \label{q2}
\end{equation}

A calculation of the radius for the top quark based on Equation (\ref{qcal}) can be found in the Appendix~\ref{A} and is $\approx 2.594\times 10^{-34}$m. We expect it to be only  a very rough estimate of the actual value of the radius. Since only one spin density is involved, the above calculation might be able to approximate the behaviour of the quarks. The calculation of the radii $r_{qx}$ for the up and down quark will probably be problematic, since their masses are not well known.

With regard to neutrinos, the self-energy function is very similar to that for charged leptons with the Z boson replacing the photon and weak coupling replacing $\alpha$.  A rough calculation gives the same order for the radius at $\approx 10^{-34}$m.  However, if this does not hold for the neutrino self-energy, then the story is quite different because they would not have self-energy due to electric or colour charge.  That~means that their rest mass-energy comes entirely from torsion energy due to intrinsic spin.  Solving that gives us a radius for electron neutrinos of the order of $10^{-26}$m.  However, the torsion (spin-squared) self-energy is negative relative to the positive rest mass.  We suspect that this means that neutrinos could have anti-gravity properties \cite{Yablon}.  The anti-gravity effect with normal matter for single neutrinos is practically negligible but the cosmological implications could be large.

\section{Possible Solution of the Hierarchy Problem}

As alluded to in the introduction, the Hierarchy Problem refers to the fact that gravitational interaction is extremely weak compared to the other known interactions in Nature. One way to appreciate this difference is by combining the Newton's gravitational constant ${G}$ with the reduced Planck's constant $\hbar$ and the speed of light ${c}$. The resulting mass scale is the Planck mass, ${m_P}$, which~some have speculated to be associated with the existence of smallest possible black holes \cite{Poplawski-2}. If we compare the Plank mass with the mass of the top quark (the heaviest known elementary particle),
\begin{align*}
m_{P} = \sqrt{\frac{\hbar c}{G}}&\approx 2.1765 \times 10^{-8} \ \textup{kg}\,, \\
m_t = \frac{173.21\ \textup{GeV}}{c^2}&\approx 3.1197 \times 10^{-25} \ \textup{kg}\,,
\end{align*}
then we see that there is some 17 orders of magnitude difference between them. This illustrates the enormous difference between the Planck scale and the electroweak scale. Many solutions have been proposed to explain this difference, such as supersymmetry and large extra dimensions, but none has been universally accepted, for one reason or another. Furthermore, recent experiments performed with the Large Hadron Collider are gradually ruling out some of these proposals. Regardless of the nature of any specific proposal, it is clear from the above values that predictions of numbers with at least 17 significant figures are necessary to successfully explain the difference between ${m_P}$ and ${m_t}$.

We saw from our numerical demonstration in the previous section that within the ECSK theory minute changes in length can induce sizable changes in the observed masses of elementary particles, and that we do have numbers at our disposal with more than 17 significant figures for producing those masses. Moreover, all length changes occurring in our demonstration are taking place close to the Planck length. Thus, since we are ``cancelling out'' near the Planck length to obtain masses down to the electroweak scale, ours is clearly a possible mechanism for resolving the Hierarchy Problem. We can appreciate this fact by simplifying our central equation by setting $\hbar = c =1$, which then reduces to
\begin{equation}
\frac{\alpha}{r_x} - \frac{3\pi G}{r_x^3} = m_x\,.
\end{equation}

It is now easy to see from this equation that the observed mass-energy only depends on the coupling constants and the radii (geometry). Moreover, it is confined entirely within a volume close to the Planck volume, as we saw in our calculations in the previous section. Thus, we are led to 
\begin{equation*}
\text{Planck Scale} \implies \text{Electroweak Scale}.
\end{equation*}

In other words, there is no hierarchy problem in the ECSK theory, because Planck scale physics is producing the electroweak scale physics in the form of the mass-energy of fermions as a byproduct of the very geometry of spacetime. 

Within the ECSK theory, which extends general relativity to include spin-induced torsion, gravitational effects near micro scales are not necessarily weak. On the other hand, since torsion is produced in the ECSK theory by the spin density of matter, it is confined to that matter, and thus is a very short range effect, unlike the infinite range effect of Einstein's gravity produced by mass-energy. In fact, the torsion field falls off as $1/r^6$, as shown in the calculations of Section \ref{III}, since it is produced by spin density squared, confined to the matter distribution \cite{Poplawski-4}.  

To compare the strengths of gravitational and torsion effects at various scales, we may define a mass-dependent dimensionless gravitational coupling constant, $Gm^2/(\hbar c)$, and evaluate it for the electron, top quark and Planck masses: 
\begin{align*}
\alpha_{G_e} = \frac{G m_e^2}{\hbar c} &\approx 1.7517 \times 10^{-45}, \\ 
\alpha_{G_t} = \frac{G m_t^2}{\hbar c} &\approx 1.1620 \times 10^{-36}, \\
\alpha_{G_P} = \frac{G m_P^2}{\hbar c} &= 1\,, \\
\alpha_e    = \frac{e^2}{4 \pi \hbar c} &\simeq 7.2973\times 10^{-3}.
\end{align*}

Here ${\alpha_e}$ is the electromagnetic coupling constant, or the fine structure constant. From these values we see that near the Planck scale the gravitational coupling is very strong compared to the electromagnetic coupling. However, as we noted above and in Section \ref{III}, near the Planck scale torsion effects due to spin density are also very strong, albeit with opposite polarity compared to that of Einstein's gravity, akin to a kind of ``anti-gravity'' effect of a very short range.

For our demonstration above we have used electrostatic energy density and spin density for matter in a static approximation, for which the field equation within the ECSK theory reduces to $G^{00} = T^{00}$. A numerical estimate for $G^{00}$ from the contributions of the electrostatic energy and spin density parts of $T^{00}$ at our cancellation-radius gives 

\begin{equation}
G^{00}_{stat} = \frac{8 \pi  G }{ c^4} \frac{\alpha \hbar c}{r_t^4}  \approx +4.209 \times 10^{62}\;\rm{m}^{-2}
\end{equation}
and
\begin{equation}
G^{00}_{spin}= -\frac{8 \pi  G }{ c^4} \frac{3\, \pi  G \hbar ^2}{c^2 r_t^6} \approx -4.209 \times 10^{62}\;\rm{m}^{-2}.
\end{equation}

Evidently, these field strengths at the cancellation radius are quite large even for a single electron. Fortunately they are never realised in Nature, because, as we can see, they cancel each other out to produce ${G^{00}_{net}=0}$. On the other hand, if we use only the mass-energy density for electron at the cancellation radius, then we obtain ${G^{00}_{mass}\approx 3.0674 \times 10^{43} \, \rm{m^{-2}}}$, which is again some 19 orders of magnitude off the mark. What is more, the latter field strength does not fall off as fast as that due to the spin-induced torsion field. Thus, it is reasonable to conclude that without the cancellation of divergent energies due to the spin self-interaction we have explored here, our universe would be highly improbable.

While there have been other approaches to the hierarchy problem from the viewpoint of the ECSK theory \cite{Zubkhov2010,Zubkhov2014,Zubkhov2013,Fabbri2016,Carillo}, our partial solution to the problem is simpler.  We have fermions with near Planck scale radii (size) producing rest mass energy in the electro-weak scale. While we have not explained why higher generation fermions do not exist, given that as an assumption, our solution is more complete.

\section{Concluding Remarks}

In this paper we have addressed two longstanding questions in particle physics: (1) Why do the elementary fermionic particles that are so far observed have such low mass-energy compared to the Planck scale? (2) What mechanical energy may be counterbalancing the divergent electrostatic and strong force energies of point-like charged fermions in the vicinity of the Planck scale? Using a hitherto unrecognised mechanism extracted from the well known Hehl--Datta equation, we have presented numerical estimates suggesting that the torsion contributions within the Einstein--Cartan--Sciama--Kibble extension of general relativity can address both of these questions in~conjunction.

The first of these problems, the Hierarchy Problem, can be traced back to the extreme weakness of gravity compared to the other forces, inducing a difference of some 17 orders of magnitude between the electroweak scale and the Planck scale. There have been many attempts to explain this huge difference, but none is simpler than our explanation based on the spin induced torsion contributions within the ECSK theory of gravity. The second problem we addressed here concerns the well known divergences of the electrostatic and strong force self-energies of point-like fermions at short distances. We have demonstrated above, numerically, that torsion contributions within the ECSK theory resolves this difficulty as well, by counterbalancing the divergent electrostatic and strong force energies close to the Planck scale.

It is widely accepted that in the standard model of particle physics charged elementary fermions acquire masses via the Higgs mechanism.  Within this mechanism, however, there is no satisfactory explanation for how the different couplings required for the fermions are produced to give the correct values of their masses. While the Higgs mechanism does bestow masses correctly to the heavy gauge bosons and a massless photon, and while our demonstration above does not furnish a fundamental explanation for the fermion masses either, we believe that what we have proposed in this paper is worthy of further research, since our proposal also offers a possible resolution of the Hierarchy~Problem.

In Reference \cite{Singh}, Singh points out that there appears to be a symmetry between small and large masses for spin-torsion coupling and energy-curvature coupling.  We have noted that there also appears to be a symmetry in that the energy-curvature coupling is effectively infinite while the spin-torsion coupling is very short-ranged near the Planck length.

One may wonder why a gravitational coupling would be involved in the torsion term involving spin-squared, but we suspect it has more to do with Planck length than gravity.  The torsion term with our $r_t$ cancellation length can be simplified to
\begin{equation}
    -\frac{3\, \pi G \hbar^2}{c^2\, r_t^3} = -\frac{3\pi (l_P)^2 \hbar c}{(\sqrt{\frac{3\pi}{\alpha}}l_P)^3} = -\frac{\alpha \hbar c}{\sqrt{\frac{3\pi}{\alpha}}l_P} \approx - 2.479 \times 10^{15}\, \textnormal{GeV}.
\end{equation}

Instead of gravitational coupling, now the term has become simple and involves only constants and the Planck length.

Needless to say, the geometrical cancellation mechanism for divergent energies we have proposed here also dispels the need for mass-renormalisation, since we have obtained finite solutions for $r_x$ taming the infinities. Thus, both classical and quantum electrodynamics appear to be more complete with torsion contributions included.  
\vspace{6pt}

\section*{Acknowledgements}
The authors wish to thank T. P. Singh for encouragement and discussions concerning the significance of torsion.

\appendix
\section{Calculations of Fermion Radii using Wolfram Mathematica}\label{A}

In this appendix we explain how we used the arbitrary-precision in {\sl Mathematica} to solve the numerical equations out to 24 significant figures.  Each equation displayed below---derived from our central Equation (\ref{result3})---is simplified so that only the numerical factors have to be used, since the dimensional units cancel out, leaving lengths in meters.  For decimal factors, the numbers must be padded out to 26 digits with zeros.  Then the numerical part of electrostatic energy density is defined as ${A}$ and the numerical part of spin energy density is defined as ${B}$, just as in Equation (\ref{A-B}) above. These~are then used throughout to perform the calculations. For the values of various physical constants involved in the calculations we have used the 2014 CODATA values, Reference \cite{CODATA} and values from the Particle Data Group, Reference \cite{PDG}.

\noindent\
\textbf{Calculation of the Cancellation Radius for Charged Leptons using Formula Equation (\ref{e}):}
\begin{equation}
{\alpha \hbar c}\,r_t^2 - \frac{3\, \pi  G \hbar ^2}{c^2}=0
\end{equation}

\noindent\(A\text{:=}N[(7.2973525664000000000000000 \times 10^{-3})(1.0545718000000000000000000 \times 10^{-34}) \\
\quad ~~~ \times (2.9979245800000000000000000 \times 10^{8}),26];\\
B\text{:=}N[(3 \pi  (6.6740800000000000000000000 \times 10^{-11}) (1.0545718001391130000000000 \times 10^{-34})^{2})/ \\
\quad ~~~ (2.9979245800000000000000000 \times 10^{8})^{2},26];\\
{N[\text{Solve}[A*r_{t}^{2} - B ==0, r_{t}], 26]\, \text{//}\text{Last}}\)\\
\noindent\(\{r_{t}\to \text{5.80838808109165274355010115 $\times$ $10^{-34}$}\}\)

\noindent\
\textbf{Calculation of Radius $\boldsymbol{r_e}$ of Electron and Positron}
\begin{equation}
\alpha \hbar c -\frac{3 \pi  G \hbar^2 }{c^2 r_{e}^2} -m_{e} c^2 r_{e} = 0
\end{equation}
\noindent\({\text{C1}\text{:=}N[(9.1093835600000000000000000 \times 10^{-31})((2.9979245800000000000000000 \times 10^{8})^{2}),26];}\\
{N\left[\text{Solve}\left[\left(A- B \left/\left (r_{e}^{2}\right)\right.\right) - \text{C1}*r_{e} ==0, r_{e}\right],26\right]\, \text{//}\text{Last}}\)\\
\noindent\(\{r_{e}\to \text{5.80838808109165274414871873 $\times {}10^{-34}$}\}\)

\noindent\
\textbf{Calculation of Radius $\boldsymbol{r_\mu}$ of Muon and Anti-Muon}
\begin{equation}
\alpha \hbar c -\frac{3 \pi  G \hbar^2}{c^2 r_{\mu}^2} -m_{\mu} c^2 r_{\mu}=0
\end{equation}
\noindent\({\text{C2}\text{:=}N[(1.8835315940000000000000000*10^{-28})((2.9979245800000000000000000*10^{8})^{2}),26];}\\
{N\left[\text{Solve}\left[\left(A- B \left/(r_{\mu}^{2})\right.\right) - \text{C2}*r_{\mu} ==0, r_{\mu} \right],26\right]\, \text{//}\text{Last}}\)\\
\noindent\(\{r_{\mu} \to \text{5.80838808109165286732522928} \times 10^{-34}\}\)

\noindent\
\textbf{Calculation of Radius $\boldsymbol{r_{\tau}}$ of Tau and Anti-Tau}
\begin{equation}
\alpha \hbar c -\frac{3\, \pi  G \hbar^2}{c^2 r_{\tau}^2} -m_{\tau} c^2 r_{\tau}=0
\end{equation}
\noindent\({\text{C3}\text{:=}N[(3.1674700000000000000000000 \times 10^{-27})((2.9979245800000000000000000*10^{8})^{2}),26];}\\
{N\left[\text{Solve}\left[\left(A- B \left/\left(r_{\tau}^{2}\right)\right.\right) - \text{C3}*r_{\tau} ==0, r_{\tau} \right],26\right]\, \text{//}\text{Last}}\)\\
\noindent\(\{r_{\tau} \to \text{5.80838808109165482503366295 $\times 10^{-34}$}\}\)

\noindent\
\textbf{Calculation of the Cancellation Radius for (2e/3) Quarks using Formula (\ref{q2}):}
\begin{equation}
\frac{9 \alpha \hbar c}{4} + \frac{\alpha_s \hbar c}{3} - \frac{3 \pi G \hbar^2}{c^{2} r_{qt}^{2}} = 0.
\end{equation}
\noindent\({\text{D}\text{:=}N[(1/3)(1/10)(1.0545718000000000000000000 \times 10^{-34})(2.9979245800000000000000000 \times 10^{8}),26];}\\
{N\left[\text{Solve}[((4/9)A+D)*r_{qt}^{2} - B ==0, r_{qt}], 26\right]\, \text{//}\text{Last}}\)\\
\noindent\(\{r_{qt}\to \text{2.59439809658779489414601733 $\times 10^{-34}$}\}\)

\noindent\
\textbf{Calculation of Radius $\boldsymbol{r_{tq}}$ of Top Quark:}
\begin{equation}
\frac{\alpha \hbar c}{6\pi} + \frac{\alpha_s \hbar c}{3}  -\frac{3 \pi  G \hbar^2} {2 c^2 r_{tq}^2} - m_{t} c^2 r_{tq}=0
\end{equation}
\noindent\({\text{E}\text{:=}N[(3.0877000000000000000000000 \times 10^{-25})((2.9979245800000000000000000 \times 10^{8})^2),26];}\\
{N[\text{Solve}[\left((4/9)A+D - B \left/\left(r_{tq}^{2}\right)\right.\right) -\text{E}*r_{tq} ==0, r_{tq}], 26]\, \text{//}\text{Last}}\)\\
\noindent\(\{r_{tq}\to \text{2.59439809658780297057798049 $\times 10^{-34}$}\}\) 

\section{Miscellaneous Derivations for S-matrix Evaluation}\label{B}

For use in deriving the following, we begin with the fermion anticommutator,
\begin{equation}
\{b_i(p), b_j^{\dag}(p')\} = \{d_i(p), d_j^{\dag}(p')\}  = (2\pi)^3 \frac{E}{m} \delta_{ij}\delta^3(\mathbf{p-p'}),\label{ac2}
\end{equation}
and since there is only one electron and no positron in both the initial and final states for the rest frame and $E_i = E_f = p_i = p_f = m = \omega = 2\pi/t$ with the spins being summed over, we have 
\begin{align}
\psi(x)\mid i\rangle &= \sqrt{\frac{m}{E_ir^3}}\psi(x)b_{i}^{\dag}(p_i)\mid 0 \rangle \nonumber \\
&= \sqrt{\frac{1}{r^3}} \int{\frac{d^3p}{(2\pi)^3}\sum_{j=1}^2\,b_j(m)u^j(m)e^{-iEt} b_{i}^{\dag}(m)\mid 0 \rangle}\nonumber \\
&=\sqrt{\frac{1}{r^3}} \int{\frac{d^3p}{(2\pi)^3}\sum_{j=1}^2\,u^{j}(m)e^{-iEt} \{b_{j}(m),b_{i}^{\dag}(m)\}\mid 0 \rangle}\nonumber \\
&=\sqrt{\frac{1}{r^3}} \,u^i(m)e^{-iEt},\nonumber
\end{align}
and similarly
\begin{align}
\langle f \mid \bar{\psi}(x) &= \sqrt{\frac{1}{r^3}} \,\bar{u}^f(m)e^{+iEt}, \nonumber \\
\psi(x)\mid i\rangle &= \sqrt{\frac{1}{r^3}} \,v^i(m)e^{+iEt},\nonumber \\
\langle f \mid \bar{\psi}(x) &= \sqrt{\frac{1}{r^3}} \,\bar{v}^f(m)e^{-iEt},\label{b5}
\end{align}
in which the fermion anticommutator Equation (\ref{ac2}) is used for simplification \cite{Milonni}. Note that the spatial components, $\sqrt{1/r^3}$, of the spin vector are factored out in these plane-wave equations.

For the derivation of Equation (\ref{ED2}), we begin with Equation (\ref{ED}) for the electrostatic self-interaction~contribution,
\begin{equation}
\textnormal{h}_I = e\,\bar{\psi}(x)\gamma^{\mu}A_{\mu}\psi(x), \label{EDrf}
\end{equation}
and substitute it into the first order S-Matrix,
\begin{equation}
S_{fi}^{(1)}(E1) = -i\,\int{d^4x\,\langle f\mid e\,\bar{\psi}(x)\gamma^{\mu}A_{\mu}\psi(x)\mid i\,\rangle}.
\end{equation}

Making substitutions using Equation (\ref{b5}) from above, and then going to the rest frame, along~with using $A_0 = e/4\pi r$, $\alpha = e^2/4\pi$ and taking $r$ to be constant, the S-matrix expression works out to give

\begin{align}
S_{fi}^{(1)}(E1) &= -i\frac{e^2}{4\pi\,r} \int{d^4\,x\, \left(\sqrt{\frac{1}{r^3}}\bar{u}^f(m)e^{+im\,t} \gamma^0\, \sqrt{\frac{1}{r^3}}u^i(m)e^{-im\,t}\right)} \nonumber \\
&= -i\frac{\alpha}{r^4}\left(\bar{u}^f(m) \gamma^0\, u^i(m)\right)\int{d^4\,x} \nonumber \\
&= -i \frac{\alpha\, r^3 t}{r^4}
\left(\begin{array}{cccc}
1& 0& 0& 0\\
0& 1& 0& 0\\
0& 0& 0& 0\\
0& 0& 0& 0
\end{array}\right) \nonumber \\
&= -i\,t\,\frac{\alpha}{r}
\left(\begin{array}{cccc}
1& 0& 0& 0\\
0& 1& 0& 0\\
0& 0& 0& 0\\
0& 0& 0& 0
\end{array}\right), \label{s-matrixED}
\end{align}
where we have taken $\int{d^4 x} = r^3\,t$.

The S-matrix evaluation for Equation (\ref{TD2}) in the rest frame is as follows. We begin with Equation~(\ref{TD}),
\begin{equation}
\textnormal{h}_I(x) = - \frac{\,3\kappa}{8}(\bar{\psi}(x)\gamma^5 \gamma_k\psi(x))(\bar{\psi}(x)\gamma^5\gamma^k\psi(x)), \label{TDHD2}
\end{equation}
and substitute it into the first order S-Matrix:
\begin{equation}
S_{fi}^{(1)}(E2) = i\frac{\,3\kappa}{8}\int{d^4x\,\langle f\mid (\bar{\psi}(x)\gamma^5\gamma_k\psi(x))(\bar{\psi}(x)\gamma^5\gamma^k\psi(x))\mid i\,\rangle}. \label{a13}
\end{equation}

Then, because the spatial components, $\sqrt{1/r^3}$, of the spin vectors are factored out in the plane-wave equations that we derived in Equation (\ref{b5}), by substituting Equation (\ref{b5}) into Equation~(\ref{a13}) and then going to the rest frame, the S-matrix expression works out to give
\begin{align}
S_{fi}^{(1)}(E2) &= i\frac{\,3\kappa}{8}\int{d^4\,x\, \frac{1}{r^3}\left(\bar{u}^f(m)e^{+imt} \,\gamma^5\gamma_0\, v^i(m)e^{+imt}\right) \,\frac{1}{r^3}\left(\bar{v}^f(m)e^{-imt}\,\gamma^5\gamma^0\, u^i(m)e^{-imt}\right)} \nonumber \\
&= i\frac{\,3\kappa}{8\,r^6}\int{d^4\,x\, \left(\bar{u}^f(m) \,\gamma^5\gamma_0\, v^i(m) \,\bar{v}^f(m)\,\gamma^5\gamma^0\, u^i(m)\right)} \nonumber \\
&= i\frac{\,3\kappa}{8\,r^6} \left(\bar{u}^f(m) \,\gamma^5\gamma_0\, (-1)
\left(\begin{array}{cccc}
0& 0& 0& 0\\
0& 0& 0& 0\\
0& 0& 1& 0\\
0& 0& 0& 1
\end{array}\right)
\,\gamma^5\gamma^0\, u^i(m)\right) \int{d^4\,x} \nonumber \\
&= i \frac{\,3\kappa\, r^3 t}{8\,r^6}
\left(\begin{array}{cccc}
1& 0& 0& 0\\
0& 1& 0& 0\\
0& 0& 0& 0\\
0& 0& 0& 0
\end{array}\right) \nonumber \\
&= i\,t\,\frac{\,3\kappa}{8\,r^3}
\left(\begin{array}{cccc}
1& 0& 0& 0\\
0& 1& 0& 0\\
0& 0& 0& 0\\
0& 0& 0& 0
\end{array}\right), \label{s-matrix0}
\end{align}
where we have again taken $\int{d^4 x} = r^3\,t$. Similarly, the S-matrix expression for the anti-fermion works out to give 
\begin{align}
S_{fi}^{(1)}(E4) &= i\frac{\,3\kappa}{8}\int{d^4\,x\, \frac{1}{r^3}\left(\bar{v}^f(m)e^{-imt} \,\gamma^5\gamma_0\, u^i(m)e^{-imt}\right) \,\frac{1}{r^3}\left(\bar{u}^f(m)e^{+imt}\,\gamma^5\gamma^0\, v^i(m)e^{+imt}\right)} \nonumber \\
&= i\frac{\,3\kappa}{8\,r^6} \left(\bar{v}^f(m) \,\gamma^5\gamma_0\,
\left(\begin{array}{cccc}
1& 0& 0& 0\\
0& 1& 0& 0\\
0& 0& 0& 0\\
0& 0& 0& 0
\end{array}\right)
\,\gamma^5\gamma^0\, v^i(m) \right)\int{d^4\,x} \nonumber \\
&= i\, \frac{\,3\kappa\, r^3\,t}{8\,r^6} (-1)
\left(\begin{array}{cccc}
0& 0& 0& 0\\
0& 0& 0& 0\\
0& 0& 1& 0\\
0& 0& 0& 1
\end{array}\right) \nonumber \\
&= -i\,t\,\frac{\,3\kappa}{8\,r^3}
\left(\begin{array}{cccc}
0& 0& 0& 0\\
0& 0& 0& 0\\
0& 0& 1& 0\\
0& 0& 0& 1
\end{array}\right). \label{s-matrix_anti}
\end{align}

From the last two equations it is easy to see that the fermion and anti-fermion S-matrix equations are coupled.  In other words, one might initially think that in the rest frame $\bar{\psi}(t)\gamma^5\gamma^0\psi(t)=0$, which~would result in the entire term being zero, but that is not the case because of the coupling of mixed states between fermions and anti-fermions is an intermediate step in the self-interaction.


\begin{thebibliography}{999}

\bibitem[Hehl(1976)]{Hehl1976}Hehl, F.W.; von der Heyde, P.; Kerlick,  G.D.; Nester, J.M.  {General relativity with spin and torsion: Foundations and prospects}. {\it Rev. Mod. Phys.} {\bf 1976}, {\it 48}, 393.

\bibitem[Trautman(2006)]{Trautman}Trautman, A. Einstein-Cartan Theory. In {\em Encyclopedia of Mathematical Physics}; Francoise, J-P., Naber, G.L., Tsun, T.S., Eds.; Elsevier: Oxford, UK, 2006; Volume 2, pp. 189--195.

\bibitem[Sciama(1964)]{Sciama}Sciama, D.W.  {The physical structure of general relativity}. {\em Rev. Mod. Phys.} {\bf 1964}, {\em 36}, 463.

\bibitem[Kibble(1961)]{Kibble}Kibble, T.W.B. {Lorentz invariance and the gravitational field.} {\em J. Math. Phys.} {\bf 1961}, {\em 2}, 212.

\bibitem[Hehl(1971)]{Hehl-Datta}Hehl, F.W.; Datta, B.K. {Nonlinear spinor equation and asymmetric connection in general relativity.} {\em J.~Math.~Phys.} {\bf 1971}, {\em 12}, 1334.

\bibitem[Poplawski(2011)]{Poplawski-1}Poplawski, N.J.  {Matter-antimatter symmetry and dark matter from torsion.} {\em Phys. Rev. D} {\bf 2011}, {\em 83}, 084033.

\bibitem[Poplawski(2012)]{Poplawski-2}Poplawski, N. J. {Nonsingular, big-bounce cosmology from spinor-torsion coupling.} {\em Phys. Rev. D} {\bf 2012}, {\em 85},~107502.

\bibitem[Poplawski(2013)]{Poplawski-3}Poplawski, N.J. {Cosmological consequences of gravity with spin and torsion.} {\em Astron. Rev.} {\bf 2013}, {\em 8}, 108.

\bibitem[Poplawski(2016)]{Poplawski-4}Poplawski, N.J. {Universe in a Black Hole in Einstein–Cartan Gravity.} {\em Astrophys. J.} {\bf 2016}, {\em 832}, 96.

\bibitem[Tecchiolli(2019)]{Tecchiolli}Tecchiolli, M. {On the Mathematics of Coframe Formalism and Einstein–Cartan Theory—A Brief Review.} {\em Universe} {\bf 2019}, {\em 5}, 206.

\bibitem[Diether(2019)]{diether2}Diether, C.F., III; Christian, J. {Existence of Matter as a Proof of the Existence of Gravitational Torsion.} {\em Prespacetime J.} {\bf 2019}, {\em 10}, 610.

\bibitem[Ortin(2004)]{Ortin}Ortin,  T. {\em Gravity and Strings: Cambridge Monographs on Mathematical Physics}; Cambridge University Press: Cambridge, UK, 2004. 

\bibitem[ohrlich(2007)]{Rohrlich}Rohrlich, F. {\em Classical Charged Particles}, 3rd ed.; World Scientific: Singapore, 2007.

\bibitem[Blagojevic(2013)]{Blagojevic}Blagojevi\'c, M.; Hehl, F.W. (Eds.) {\em Gauge Theories of Gravitation}; World Scientific: Singapore, 2013.

\bibitem[Freidel(2012)]{Freidel}de Berredo-Peixoto, G.; Freidel, L.; Shapiro, I.L.; de Souza, C.A. {Dirac fields, torsion and Barbero-Immirzi parameter in cosmology.} {\em J. Cosmol. Astropart. Phys.} {\bf 2012}, {\em 6}, 017.

\bibitem[Magueijo(2013)]{Magueijo}Magueijo, J.; Zlosnik, T.G.; Kibble, T.W.B.  {Cosmology with a Spin.} {\em Phys. Rev. D} {\bf 2013}, {\em 87}, 063504.

\bibitem[Rudenko(2014)]{Rudenko}Rudenko, A.S.; Khriplovich, I.B. {Gravitational four-fermion interaction in the early Universe.} {\em Phys. Uspekhi} {\bf 2014}, {\em 57}, 167.

\bibitem[Boos(2017)]{Boos}Boos, J.; Hehl, F.W. {Gravity-induced four-fermion contact interaction: liberating the intermediate W and Z type gauge bosons.}, {\em Int. J. Theor. Phys.} {\bf 2017}, {\em 56}, 751.

\bibitem[Fabbri(2011)]{fabbri}Fabbri, L.; Vignolo, S. {Dirac fields in f(R)-gravity with torsion.} {\em Classical Quantum Gravity} {\bf 2011}, {\em 28}, 12.

\bibitem[Weisskopf(1939)]{Weisskopf}Weisskopf, V.F. {On the Self-Energy and the Electromagnetic Field of the Electron.} {\em Phys. Rev.} {\bf 1939}, {\em 56}, 72.

\bibitem[Milonni(1994)]{Milonni}Milonni, P.W. {\em The Quantum Vacuum: An Introduction to Quantum Electrodynamics}; Academic Press: San Diego, CA, USA, 1994; Chapter 12.

\bibitem[Weinberg(2005)]{Weinberg}Weinberg, S. {\em The Quantum Theory of Fields, Vol. I}; Cambridge University Press, New York, New York, USA, 2005; Chapter 11.

\bibitem[Halzen(1984)]{Halzen}Halzen, F.; Martin, A. {\em Quarks and Leptons: An Introductory Course in Modern Particle Physics}; Wiley, Inc.: Hoboken, NJ, USA, 1984; Chapter 7.

\bibitem[Dehmelt(1988)]{Dehmelt}Dehmelt, H. {A Single Atomic Particle Forever Floating at Rest in Free Space: New Value for Electron Radius.} {\em Phys. Scr.} {\bf 1988}, {\em 1988-T22}, 102.
\bibitem[Poplawski(2017)]{Poplawski-5}Poplawski, N.J. {Uncertainty principle for momentum, torsional regularization, and bare charge.} {\em arXiv} {\bf 2017}, arXiv:1712.09997.

\bibitem[Poplawski(2018)]{Poplawski-6}Poplawski, N.J. {Torsional regularization of vertex function.} {\em arXiv} {\bf 2018}, arXiv:1807.07068.

\bibitem[Yablon(2013)]{Yablon}Yablon, J. {Grand Unified SU(8) Gauge Theory Based on Baryons which Are Yang-Mills Magnetic Monopoles.} {\em J. Mod. Phys.} {\bf 2013}, {\em 4}, 94.

\bibitem[Zubkhov-a(2010)]{Zubkhov2010} Zubkhov, M.A. Torsion Instead of Technicolor. {\em Mod. Phys. Lett. A} {\bf 2010}, {\em 25}, 2885.

\bibitem[Zubkhov-b(2014)]{Zubkhov2014} Zubkhov, M.A. Dynamical torsion as the microscopic origin of the neutrino seesaw. {\em Mod. Phys. Lett. A} {\bf 2014}, {\em 29}, 1450111.

\bibitem[Zubkhov(2013)]{Zubkhov2013} Zubkhov, M.A. Gauge theory of Lorentz group as a source of the dynamical electroweak symmetry breaking. {\em J. High Energy Phys.} {\bf 2013}, 44. doi:10.1007/JHEP09(2013)044.

\bibitem[Fabbri(2016)]{Fabbri2016} Fabbri, L. A simple assessment on the hierarchy problem. {\em Int. J. Geom. Meth. Mod. Phys.} {\bf 2016}, {\em 13},  1650068.

\bibitem[Castillo(2013)]{Carillo} Castillo-Felisola, O.; Corral, C.; Villavicencio, C.; Zerwekh, A.R. Fermion masses through condensation in spacetimes with torsion. {\em Phys. Rev. D} {\bf 2013}, {\em 88}, 124022.

\bibitem[Singh(2018)]{Singh}Singh, T.P. {A new length scale, and modified Einstein-Cartan-Dirac equations for a point mass.} {\em Int. J. Mod.~Phys.} {\bf 2018}, {\em 27}, 1850077.

\bibitem[CODATA(2016)]{CODATA}Mohr, P.J.; Newell, D.B.; Taylor, B.N. {CODATA Recommended Values of the Fundamental Physical Constants: 2014.} {\em Rev. Mod. Phys.} {\bf 2016}, {\em 88}, 035009.

\bibitem[PDG(2014)]{PDG}Olive, K.A.; [Particle Data Group] {et al.} Particle Data Group. {\em Chin. Phys. C} {\bf 2014}, {\em 38}, 090001; see also 2015 update.

\end{thebibliography}
\end{document}